# Static and hydrodynamic periodic structures induced by AC electric fields in the antiferroelectric SmZ$_A$ phase

K. S. Krishnamurthy,[1,*] S. Krishna Prasad,[1] D. S. Shankar Rao,[1] R. J. Mandle,[2] C. J. Gibb,[3] J. Hobbs[4] and N. V. Madhusudana[5,*]

[1]*Centre for Nano and Soft Matter Sciences, Bangalore 562162, India*

[2]*School of Physics and Astronomy, School of Chemistry, University of Leeds, Leeds LS2 9JT, UK*

[3]*School of Chemistry, University of Leeds, Leeds LS2 9JT, UK*

[4]*School of Physics and Astronomy, University of Leeds, Leeds LS2 9JT, UK*

[5]*Raman research Institute, Bangalore, 560080, India*

## ABSTRACT

We report the effect of AC electric fields in the range of 0.1-300 kHz on planar antiferroelectric SmZ$_A$ layers of DIO. Significant results are (a) primary bifurcation into a quasistationary periodic instability with its voltage threshold $U_c$ and wave vector $\mathbf{q}_c$ *along* the initial director being, respectively, quadratic and linear functions of $f$ over 10-150 kHz, and with an azimuthal distortion of the director which changes sign between adjacent stripes, (b) transition from the modulated planar state to a homogeneous state at higher voltages, and (c) third bifurcation into travelling wave periodic state on further rise in $U$ in the region 10-40 kHz. We interpret these findings as follows. The low voltage instability is very similar to that seen in the higher temperature apolar nematic phase, and is the electrohydrodynamic (EHD) instability possibly belonging to the region of dielectric inversion frequency. The azimuthal distortions of $\mathbf{n}$ result from an undulatory distortion of the SmZ$_A$ layers in the book-shelf geometry. The intermediate homogeneous state of SmZ$_A$ in which the periodic structure is absent results from a linear coupling between the layer polarization $\mathbf{P}$ and applied field $\mathbf{E}$, giving rise to a scissoring type mutual P reorientation in adjacent layers. Finally, at even higher voltages, the medium goes over to a field induced transition to the ferroelectric nematic, with the polarization following the AC field, and the periodic EHD instability being similar to that of the dielectric regime. The polar vector symmetry of the medium leads in general to travelling waves.

*murthyksk@cens.res.in

# 1. INTRODUCTION

Nematic (N) liquid crystals are characterised by an *apolar* director (**n**) [1]. Even when the constituent rod-like molecules have strong longitudinal dipoles, near-neighbouring molecules prefer an antiparallel orientation favoured by electrostatic interaction, ensuring that **n** remains apolar [2]. Such compounds, by their relatively large positive dielectric anisotropy $\Delta\varepsilon=\varepsilon_\parallel\text{-}\varepsilon_\perp$ and lowered reorientational thresholds, have become indispensable to LCDs, which have dominated the display industry during the past few decades.

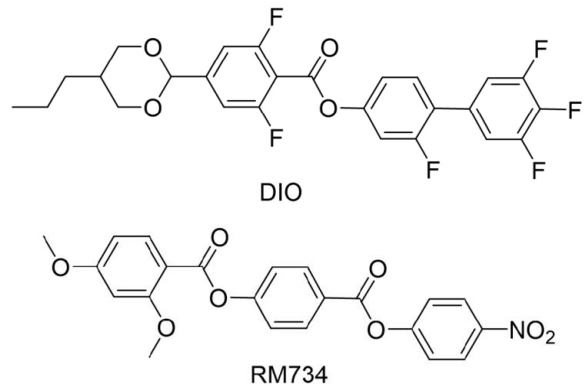

FIG. 1. Chemical structures of DIO [(2,3',4',5'-tetrafluoro-[1,1'-biphenyl]-4-yl 2,6-difluoro-4 (5-propyl-1,3-dioxane-2-yl) benzoate] and RM734 [4-[(4-nitrophenoxy)carbonyl]phenyl 2,4-dimethoxybenzoate] that exhibit the ferroelectric nematic order.

Remarkably, two new rod-like molecules (Fig. 1) with very large longitudinal dipole moments (> 9D) have been reported recently [3-5] as exhibiting new nematic-to-nematic transitions, the nematic at the lowest temperatures being identified as ferroelectric ($N_F$) with a *polar* symmetry [6, 7]. The polarization **P** is found to saturate at a rather high value, exceeding 6 µC/cm². In the past couple of years, more than 100 ferro-nematogens have been synthesised, and a variety of physical studies reported on many of them [8,9]. Though Max Born developed the first molecular theory of nematics based on dipolar intermolecular interactions, suggesting it to be a ferroelectric medium [10], as mentioned in the previous paragraph a typical nematic with polar molecules exhibits apolar symmetry. As one of us [11] has argued, the $N_F$ phase is exhibited by compounds with a *specific* structure of their rod-like molecules. The latter have longitudinal surface charge density waves which build up the large dipole moments, with the requirement that the amplitudes of the half waves at the two ends being significantly lower than those in the middle. By and large all the ferro-nematogens synthesised so far conform to this requirement [9].

Electric field effects in the N phase have long been a subject of great technological and scientific interest [12]. The most widely known of these is the Fréedericksz transition involving a homogeneous reorientation that enables measurement of curvature elastic constants; it forms the basis of operation in nematic LCDs. The destabilising torque on the director arises due to the dielectric anisotropy $\Delta\varepsilon$. As the frequency of the applied voltage is increased, $\varepsilon_\parallel$ can relax at a relatively low frequency, in some cases leading to an inversion in the sign of $\Delta\varepsilon$. Periodic deformations have been found around this frequency as well [13-17]. The N phase also exhibits electrical conductivity anisotropy $\Delta\sigma=\sigma_\parallel-\sigma_\perp$ of dissolved ionic species. Nematics with negative $\Delta\varepsilon$ and positive $\Delta\sigma$ [(− +)] exhibit an electrohydrodynamic instability (EHD) in planar samples above a threshold voltage due to the well-known Carr-Helfrich mechanism [18], in which thermal fluctuations with a bend curvature of the **n** field can generate a space charge density which is subject to forces due to the electric field. It results in a periodic structure involving flow of the nematic. A complete analysis of the instability requires a knowledge of all the physical properties of the medium [19, 20]. At low frequencies of an applied AC electric field, the charge densities can follow the electric field **E**, leading to the conduction regime. The charges relax at a frequency $f_c$, beyond which the curvature of the director itself can oscillate with **E**, with the wavelength of the periodic instability shrunk to a low value [1]. A full analysis including proper boundary conditions is quite involved, and has been accomplished, and it compares very favourably with experimental results [1,21,22].

If the nematic sample has splay or bend distortions of the **n** field, it exhibits 'flexoelectric' polarization [23], quite generally as the N phase has quadrupolar order [24]. In general, at low frequencies, flexoelectric effects give rise to oblique EHD rolls [25,26], and also electroconvection just above smectic to nematic transition point, if both $\Delta\varepsilon$ and $\Delta\sigma$ have negative signs and the Carr-Helfrich mechanism is inapplicable [25]. Interesting types of EHD patterns are also found in nematics made of bent-core (BC) molecules [27-30], which are known to be made of clusters of ~100 molecules [31] giving rise to a sequence of relaxational responses.

The intrinsic polarization **P** of the $N_F$ phase results in a *linear* coupling with **E**. Many types of defects in the **P** field generate charge densities and hence force densities, leading to electroconvection under the action of **E**. In sandwich cells, **P** prefers to avoid any component normal to the cell surfaces. An **E** field applied between the surfaces results in a block reorientation of **P** depositing charges at the surfaces, i.e., the medium acts like a conductor, and as recent results have shown [32-34], the interpretation of the large capacitances measured in such cells is quite delicate.

Of the two compounds in Fig. 1, DIO is particularly interesting as it is found to exhibit a third phase in between the N and $N_F$ phases [3]. In some compounds, the intermediate antiferroelectric phase, denoted as $N_S$ [35], exhibits a microscale modulation in a certain temperature range, which is attributed to a periodic splay distortion of the P field; the origin of $N_S$ has been traced to flexoelectric effect [7]. The period of modulation appears to depend on the sample thickness. The $N_S$ phase with a layer spacing in nm range, which is not dependent on sample thickness, probably has a different physical origin. Detailed physical studies by Chen et al. [36] have established that the medium has a *layered* structure, with a spacing of ~20 nm consisting of a pair of antiferroelctric polarized domains, **P,** and hence the long axes of the molecules are orthogonal to the layer normal. Thus, while the higher temperature N phase has antiparallel orientations of near neighbouring molecules, the intermediate $N_s$ phase, referred as $SmZ_A$ in [36], has antiferroelectric order with a supramolecular domain structure, limited along the layer normal to a few nanometres (in DIO), ensuring the net **P** of a macroscopic sample to be zero, as in the N phase. If a sufficiently large **E** field is applied to the sample, **P** aligns parallel to **E** in all the domains, and a uniform field induced $N_F$ phase results. Nacke et al. have used a phenomenological model proposed to describe a similar sequence of phases exhibited by some solid crystals to analyse the experimental results. Recently, several compounds have been synthesised which exhibit a similar sequence with an intermediate $SmZ_A$ phase occurring between the higher temperature N and lower temperature $N_F$ phases [37].

Though the $SmZ_A$ phase is *biaxial* unlike the N phase, the antiparallel packing in both of them at relatively small length scales has prompted us to investigate the **E** field induced distorted structures in both, and look for similarities as well as differences in their responses, using the prototype compound DIO. It is relevant to mention here of a recent report on the electric instabilities in $SmZ_A$ layers of DIO, held in a 5 μm cell treated for planar alignment [38]; it concerns the longitudinal (parallel to the rubbing direction R) and transverse striped states excited at a fixed frequency of 1 kHz, under varying voltages and temperatures. These states have been interpreted by the authors as corresponding to *equilibrium* structures with different types of periodic splay-bend deformation of the **P**-field proposed in their theoretical model; the transformations under the electric field are suggested to be due to a complex interplay between flexoelectric and induced polarizations. The present report deals with electric instabilities in initially planar DIO layers of varying thickness (5, 9 and 20 μm) at different temperatures; the response of the medium is examined at different voltages, over a wide range of frequencies (0.1-300 kHz). While our main focus is on the antiferroelectric phase, for comparative analysis, we have also studied the instabilities in the para- and ferroelectric nematic phases. We present the static and dynamical features of the various states and point to some novel results in the $SmZ_A$ phase. We discuss the results in terms of some plausible reorientational and electrohydrodynamic processes.

## II. EXPERIMENTS

The compound DIO is known to exist in different isomeric forms [39,40] among which are the non-mesogenic *cis* and mesogenic *trans*. The latter, on cooling the liquid, goes through the sequence Isotropic I–paranematic N–antiferroelectric smectic $SmZ_A$–ferroelectric nematic $N_F$–crystal Cr. For the sample of DIO used here the phase sequence was I (170 ºC) N (82 ºC) $SmZ_A$ (68 ºC) $N_F$. Optical textures are from a Carl-Zeiss Axio Imager.M1m polarizing microscope having the provision for recording time-lapse and z-stacked images through an AxioCam MRc5 digital camera. For recording fast dynamical changes, a Sony camera (A7Cii) was used, with a frame rate of 50 fps, and the extracted stills were analysed using ImageJ. The sample cells, procured from M/s Instec and M/s AWAT, were sandwich type, constructed of ITO coated glass plates with overlaid polyimide layers. The plates buffed unidirectionally along *R*, and stated to provide a tilt angle of less than 3º, were assembled so that *R* remained antiparallel at the two limiting surfaces; sample thickness *d* varied between 5 μm and 20 μm. In POM studies, an Instec HCS402 hot-stage coupled to a STC200 controller was employed to maintain the sample temperature *T* to an accuracy of ±0.1˚C. For electric field experiments, a Stanford Research Systems function generator (DS345) linked to a FLC Electronics voltage amplifier (model A800) was used. The frequency *f* and rms voltage *U* of the external field were recorded with a Keithley-2002 multimeter. Permittivity data were obtained using a precision LCR meter (Agilent E4980A) capable of providing up to 20 $V_{rms}$; the probing frequency was 1 kHz, unless stated otherwise. An FP82HT hot-stage coupled to a FP90 controller with a ±0.1ºC precision was used in all temperature dependent capacitance measurements; the probing voltage was kept low (10 mV) in order to be well below any reorienting field. Dielectric measurements were initiated by heating the sample to an initial temperature of 95 ºC, and subsequently bringing it down to the required value. For convenience, we use the following notations: the orthogonal reference axes *x* and *z* define the alignment and observation/electric field directions, respectively; the rubbing direction R in an antiparallel rubbed cell lies along +*x* at one substrate and along –*x* at the other. P(*α*)–A(*β*) indicates the setting of the polarizer P and analyser A with their axes at angles *α* and *β* (degrees) relative to *x*; see Fig. 2 for experimental geometry.

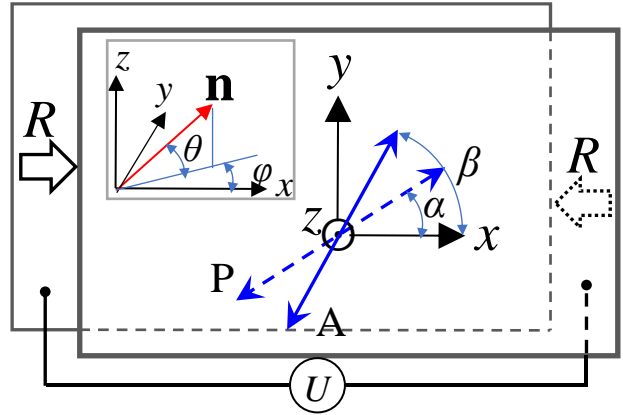

FIG. 2. Experimental geometry showing a planar aligning cell with antiparallel rubbing along R at the substrates. Polarizer P and analyzer A are with their transmission axes at angles $\alpha$ and $\beta$ relative to *x*, respectively. The initial director $\mathbf{n}_o$ in the N phase lies along *x*. Light traverses along *z*, toward the observer. The inset shows the conventional definition of polar and azimuthal angles $\theta$ and $\varphi$, respectively.

## III. RESULTS

### A. Instabilities in the antiferroelectric phase

We may first consider the observations made in the mid temperature region of the SmZ$_A$ phase, at a relative temperature $T_r = T_{AF} - T \approx 5$ °C, where $T_{AF}$ denotes the onset temperature of the antiferroelectric phase. In the frequency region of 100 Hz to 4 kHz, the initial planar state is destabilized by dielectric reorientation at a threshold that increases with frequency from about 200 to 450 mV. The first bifurcation of the planar state into a *regularly* modulated periodic state, with the wave vector **q** parallel to the rubbing direction R, is obtained across a wide frequency region of 5 kHz to 300 kHz. For illustrating the typical morphological changes, we consider the results obtained at a fixed *f* of 10 kHz. Close to the voltage threshold, $U_c$, the instability is localized, appearing often as chains of rectangular domains along R as in Figs. 3(a) and 3(b), and developing into an extended pattern with time, as in Fig. 3(c). The latter process may be hastened using a slightly higher voltage $U > U_c$. The stripes are very clearly seen when a single polariser is oriented parallel to wavevector **q**, as evident in Fig. 3(e). This indicates that the main distortion of the sign from one stripe to the next. The appearance of well-defined focal lines in (e) for P(0) and lowered focusing effect in (f) for P(90) indicate the focal lines as predominantly due to out-of-layer-plane deviation $\theta$ of the director.

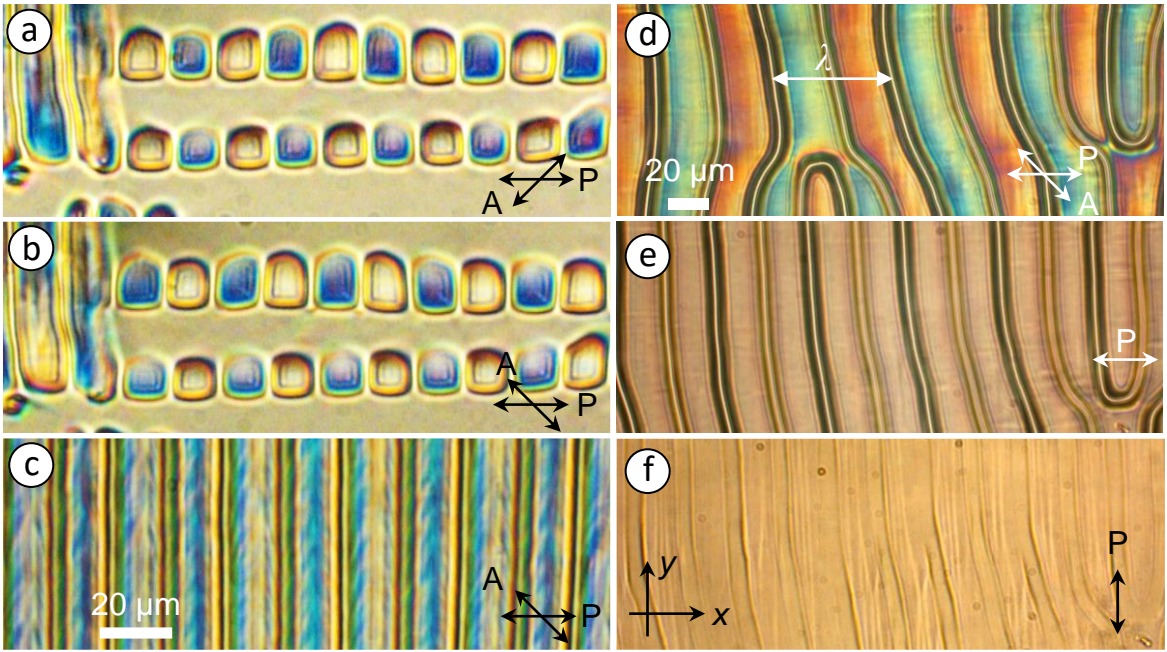

FIG. 3. (a, b) A 9 μm thick layer of DIO at 75 °C subjected to a 0.8 V, 10 kHz field showing rectangular domains of distortion soon after application of the field; (c) appearance of the extended region of domains with the wavevector **q** along R (or *x*), developed over several minutes. (d-f) A 20 μm thick layer of DIO at 75 °C subjected to a 1 V, 10 kHz field. In (d), *λ* is the pattern period; alternating colors of successive stripes arise from the azimuthal deviation changing

director is in the plane containing **E** and R, the rubbing direction along which the undistorted director $\mathbf{n}_o$ aligns. The pattern is described by a periodic spatial variation of the polar angle $\theta$ that leads to the formation of an array of equidistant focal lines parallel to $y$ when light vibrating along R traverses the sample. When using a single polarizer (either P or A) with its axis along $y$, the pattern contrast greatly diminishes, but the focal lines still remain discernible, as in Fig. 3(f). In addition to its periodic deviation in the $xz$ plane, therefore, the director also possesses a significant component in the $xy$ plane, which is described by the azimuthal angle $\varphi$ that changes sign from one stripe to the next; compare Figs. 2(a) and 2(b). Remarkably, while such a nonzero $\varphi$ angle (which, for example, arises in the Bobylev-Pikin flexoelectric effect [41]) leads to a nonzero angle between **q** and $\mathbf{n_o}$ [42], the latter angle is *zero* in the present system. The director field in the periodic state represented by Fig. 3 that involves both polar and azimuthal director deviations may be visualized as in Fig. 4(a); it shows schematically the **n**-field in the sample midplane ($z$=0) with nail heads therein indicating out-of-plane tilts; D and B refer to dark and bright bands along $y$ for P(0)-A(90). Periodic textures in a 20 μm thick planar layer at 75 ℃, driven by a 10 kHz, 1.1 V field are in agreement with this model; for example, the focal lines in Fig. 4(c) are located along the dark bands in Fig. 4(b) that outline the vertical sections through the undisturbed director.

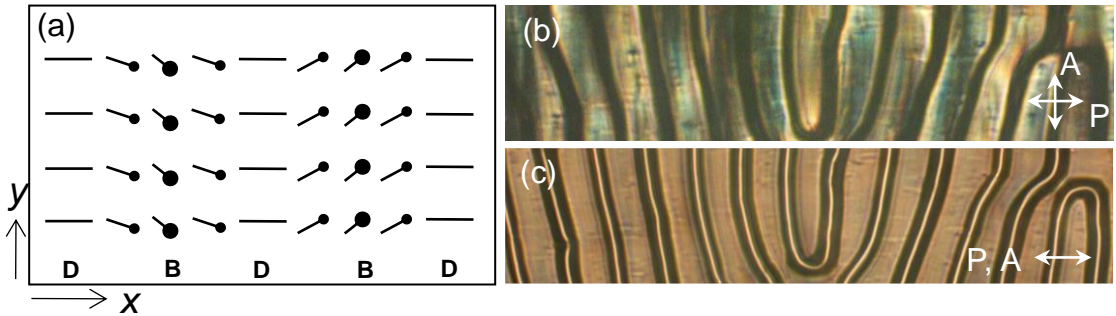

FIG. 4. Schematic of the **n**-field in the sample midplane ($z$=0). Nail heads indicate out-of-plane tilts. D and B refer to dark and bright bands along $y$ for P(0)-A(90). (b,c) Periodic textures in a 20 μm thick planar layer at 75 ℃, driven by a 10 kHz, 1.1 V field. In conformity with the model in (a), the focal lines in (c) appear along the dark bands in (b) that delineate the vertical sections of undisturbed director.

The periodic state has a voltage threshold $U_c$ (rather than a field threshold) since its value remains approximately the same for different thicknesses $d$, at a given frequency and temperature. $U_c$ is strongly dependent on frequency; as Fig. 5 shows, it scales quadratically with $f$ up to about 180 kHz and thereafter the rise slows down rapidly, showing a change in curvature. The pattern period period also depends strongly on frequency, decreasing exponentially with increasing $f$, as seen in Figs. 6 and 7. The magnitude of critical wave number $q_c$ varies linearly up to about 150 kHz and saturates thereafter. The linear segment extends until the pattern wave length drops slightly below the sample thickness. We may note that this type of strong dependence of $\lambda$ on $f$ in the high frequency regime is reminiscent of a similar occurrence in bent-core nematics [28]. The longitudinal and transverse inplane rolls (meaning that the vortices of flow lie in the layer plane) of electroconvection therein also show the variable grating effect with respect to frequency.

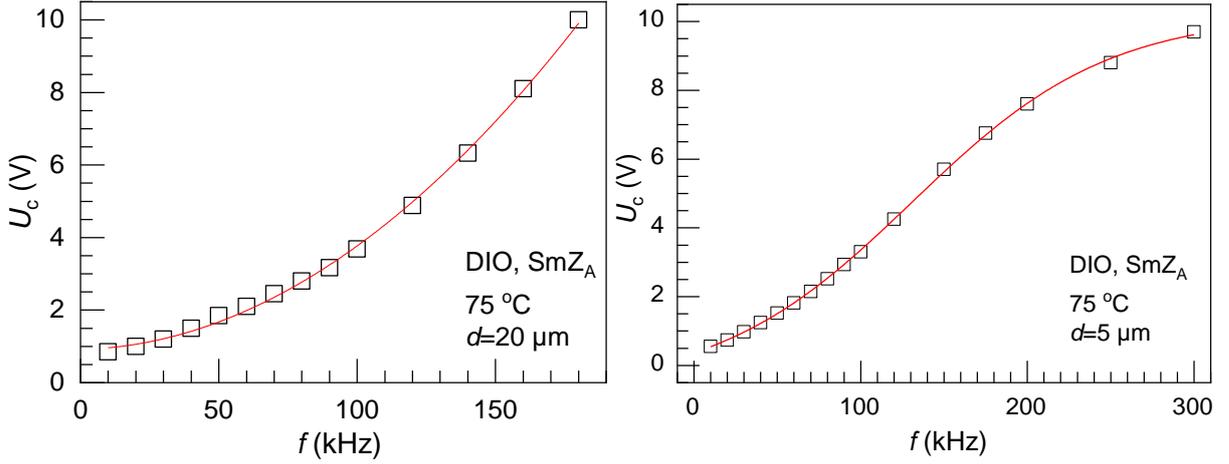

FIG. 5. Variation of the critical voltage $U_c$ at which the first bifurcation into the striped state occurs as a function of frequency $f$; d=20 µm (left), 5 µm (right). Red lines are (left) polynomial and (right) Boltzmann model fits.

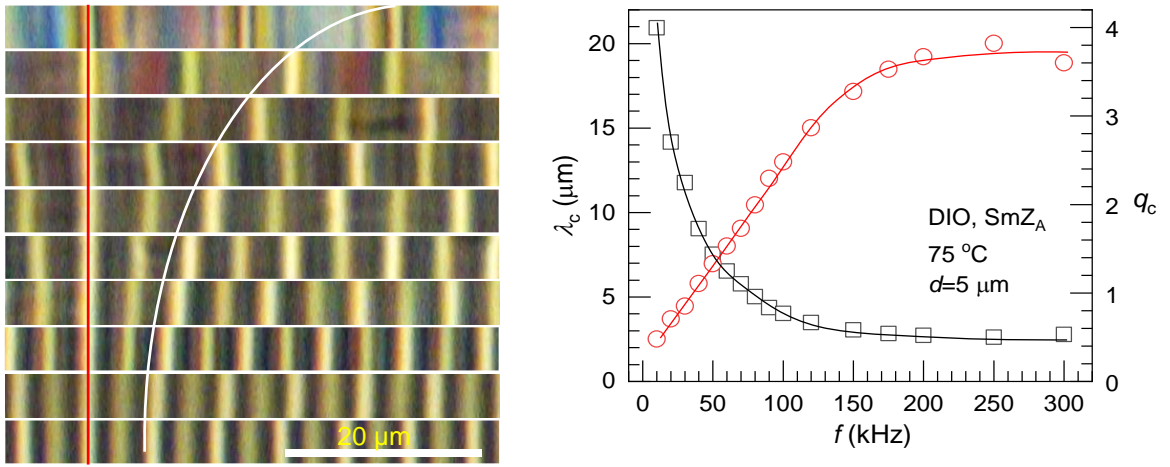

FIG. 6. (left) Reducing pattern period with increasing frequency in a 9 µm thick sample of DIO at 75 ºC. The frequency is increasing from top to bottom in steps of 10 kHz, starting at 10 kHz at the top. The patterns are recorder at voltages close to threshold. The red line is for reference. The curved line outlines the location of stripes one period from reference.

FIG 7. (Right) Variation with frequency $f$ of threshold period $\lambda_c$ and wave number $q_c$ in units of $\pi$/d in a 5 µm thick sample of DIO at 75 ºC. Continuous lines are guides to the eye.

At low frequencies, the distortion continues to increase with voltage up to a critical value after which the sample loses the periodic structure altogether and assumes a *uniform* appearance. The discontinuous manner in which this comes about is characteristic of the transition, which is illustrated in Fig. 8. In Fig. 8(a), the stripes display vivid colours determined by periodic ($\theta$, $\varphi$) deviations of the director. Upon increasing the voltage from 1 V to 2 V, as in Fig. 8(b), the colors tend toward grey, pointing to lowering birefringence. The boundaries between adjacent bands acting as Brochard-Leger walls [43,44] get narrowed and dissociate into pairs of opposite wedge lines. The advancing train of disclinations leads to the first order transformation of the stripe state into the uniform state. The birefringence of the uniform state is much lowered compared to that of the field-free base state. For example, using a tilt-compensator, we found the optical path of a 20±2 μm sample at 75 °C to reduce from 4280 nm in the base state to 1520 nm in the homogeneous state derived using a 1.45 V, 10 kHz field.

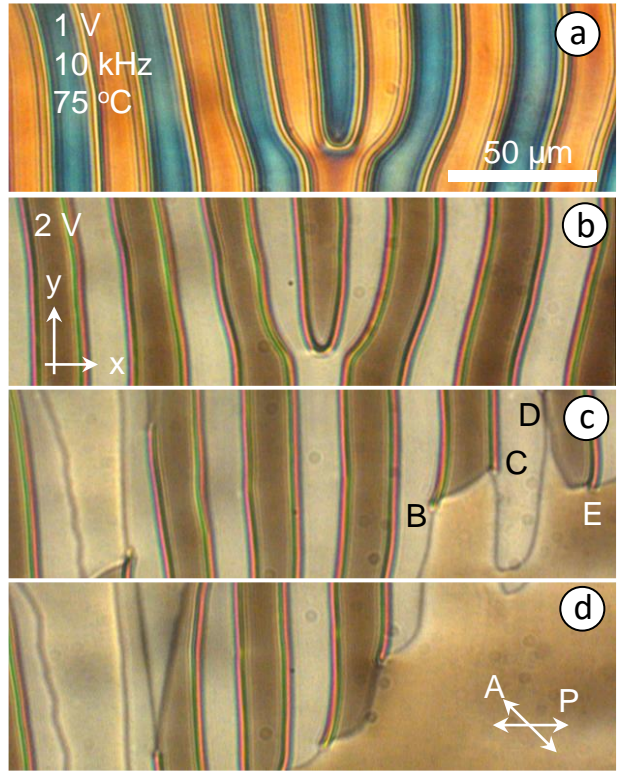

FIG. 8. Transition from the periodic state to homogeneous state mediated by the pincement process along the domain boundaries acting as Brochard-Leger walls. In (a), $U$ is sufficiently low for the birefringence colours to be vivid; on doubling $U$, lowered effective birefringence shifts the colors toward grey, as in (b-d); and, with passage of time, the wall-like boundary regions along focal lines undergo pincement or dissociation into a pair of opposite wedge lines of strength s=|1/2|; for example, at each of the junctions B-E a pair of opposite lines has formed. As the pincement progresses, the homogeneous state enlarges at the expense of the patterned state.

As the voltage is increased further in the frequency range up to about 40 kHz, a second periodic instability develops from the homogeneous state. It is characterised by the travelling wave (TW) phenomenon right from its onset (Hopf bifurcation) [22]. In the initial stage of evolution of the TW pattern under a slowly increasing $U$, finger-like stripes of limited extension along $y$ form sporadically; they are always found in a dynamical state, traveling along $x$ or -$x$. With a marginal increase in $U$, they extend to fill the entire sample. When the voltage is rapidly increased from a low value, the static striped state goes over discontinuously to the TW state with the intermediate homogeneous state not revealing itself, as illustrated in Fig. 9. As in the usual TW state of calamitics, the waves propagate in opposite directions in different parts of the sample; see Figs. 9(d)-9(g). The **q** vector of the TW instability is generally confined to a narrow angle with respect to **n₀**. A typical pattern of TWs observed in a 5 µm thick sample at 75 ºC under a 2.2 V, 10 kHz field, with a single polarizer set along $x$, is shown in Fig. 10(a). We used a Sony camera (A7Cii) for video recording the TWs at 50 fps. From the intensity profiles along a chosen line (parallel to R) in a succession of extracted still images with equal time intervals, we obtained the TW speed $v_{TW}$ and the Hopf frequency $f_{TW}=v_{TW}/\lambda$, as noted in the caption to Fig. 10. With increase in applied voltage, both the velocity $v_{TW}$ and frequency $f_{TW}$ of TWs are found to vary linearly regardless of sample thickness; see Fig. 11. With increasing temperature, both the critical voltage and velocity of the traveling waves increase exponentially; see Fig. 12.

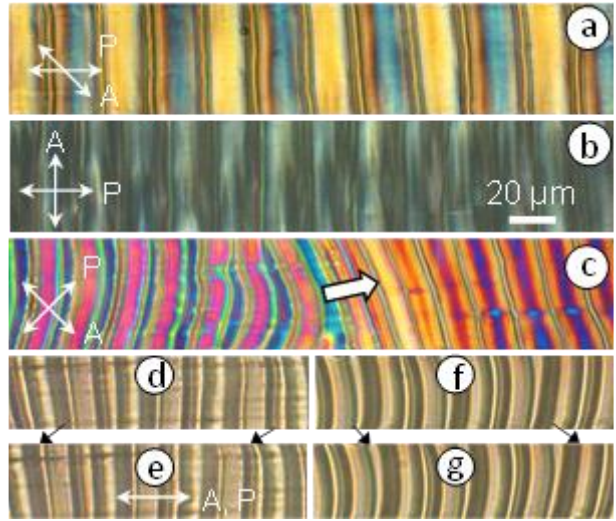

FIG. 9. (a-c) Transition from modulated planar to modulated homeotropic state upon a sudden, large rise in *U*; *d*=20 μm. (a) Soon after voltage jump from 1 V to 2 V; domains in the planar geometry. (b) Same as (a), under crossed polarizers. (c) Periodic state of the quasihomeotropic region invading the modulated planar region from left. (d-g) Pattern in the quasihomeotropic state propagating at 2 V in opposite directions; drift is to the left in (d, e), and to the right in (f, g); sloping arrows indicate the extent of domain displacement in 0.26 s that separate (d) and (e), as also (f) and (g).

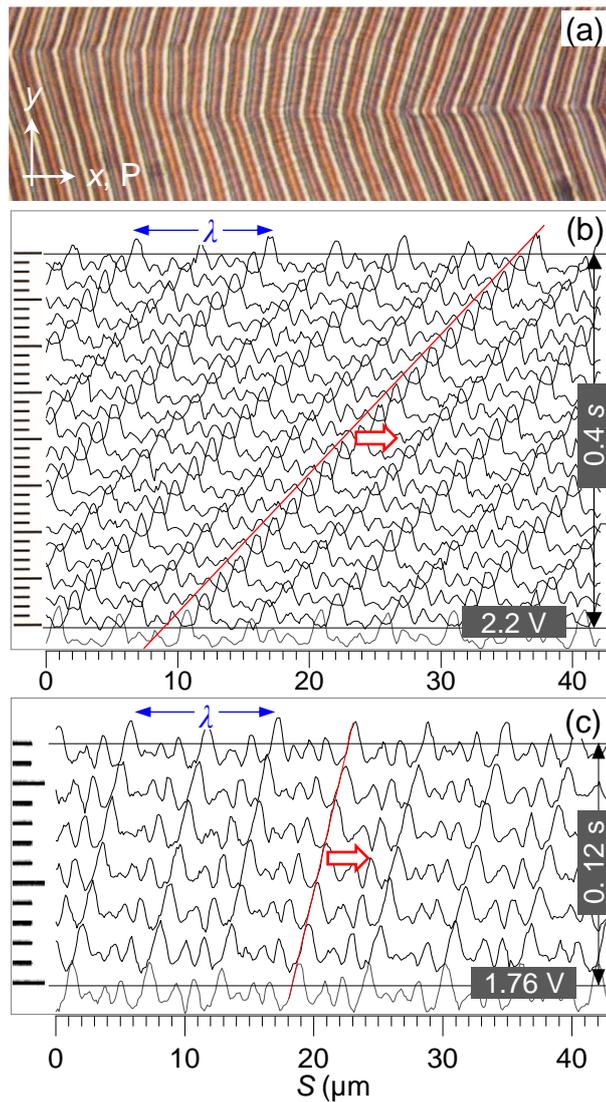

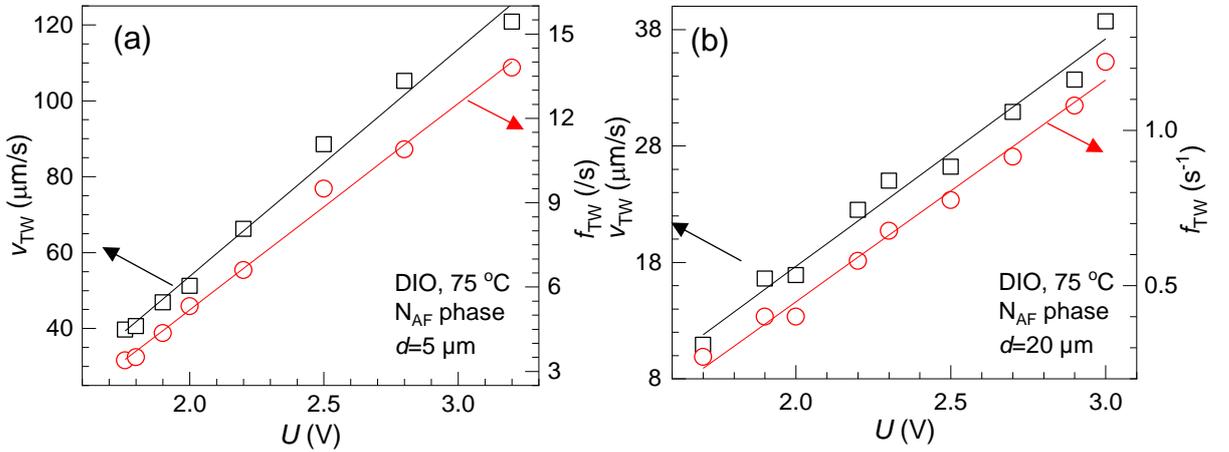

FIG. 11. Linear variation of both the velocity and frequency of TWs with the applied voltage (10 kHz), for thickness $d$=0.5 µm (a) and 20 µm (b).

Above ~40 kHz, the dynamical features get substantially modified. For example, in a 20 µm thick sample at 50 kHz, the stripes normal to **n₀** and stationary at 1.8 V become undulatory (along $y$) at higher $U$. Simultaneously, edge dislocations nucleate randomly at the Zig and Zag turnings and drift rapidly along $x$ or -$x$, while the stripes between the lines defined by the zig-zag junctions remain nearly stationary. At 2.7 V, initially, the uniform and periodic states dynamically exist until the patterned state with reduced wavelength is finally established. At even higher voltages, hydrodynamic patterns of increasing complexity appear; see Figs. 13(c) and 13(d). Beyond about 200 kHz, near the threshold, the stripes become undulatory along $y$ and $q$ tends to saturate. We need to note that patterns realized at higher voltages and frequencies are difficult to assign to a particular temperature because of dielectric heating. Their interpretation would require measuring $T$ right at the sample site.

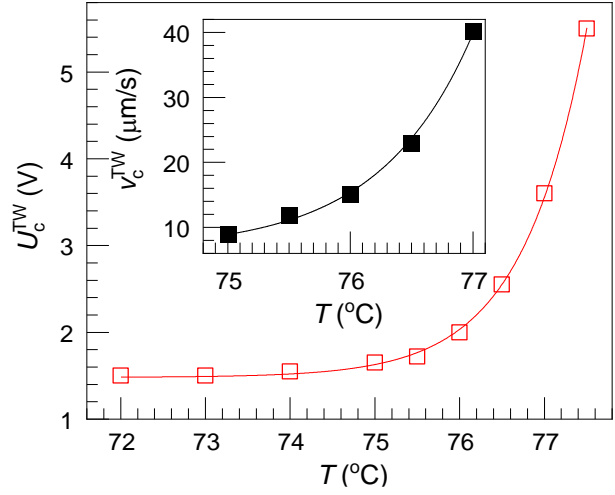

FIG. 12. Exponential increase of the TW critical voltage with temperature. *Inset*: Exponential increase of threshold TW velocity with temperature. Continuous curves are exponential fits. $d$=20 μm, $f$= 10 kHz.

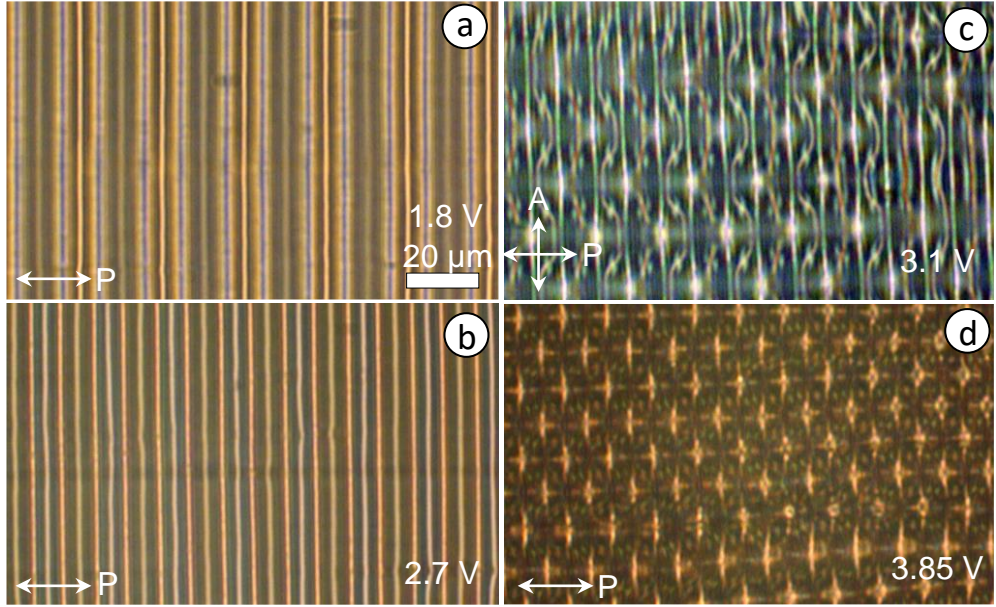

FIG. 13. Instabilities obtained at 50 kHz in a 20 μm thick DIO layer in the SmZ$_A$ phase at 75 ℃. Soon after rising $U$ to 2.7 V from 1.8 V, periodic and uniform states dynamically exist for a while after which the pattern with lowered period in (b) ensues. Complex dynamical patterned states appear at higher voltages (c,d).

## B. Instabilities in the paranematic phase

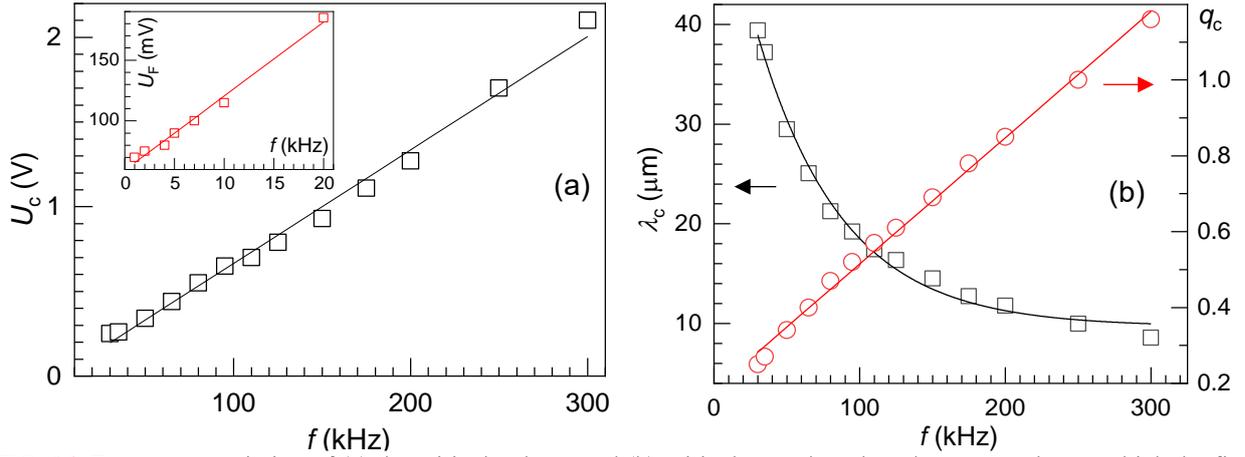

FIG. 14. Frequency variation of (a) the critical voltage and (b) critical wave length and wave number at which the first bifurcation into the striped state occurs in a 5 μm thick DIO layer in the paranematic phase, at 5 °C above the SmZ$_A$-N transition point. Inset to (a): Freedericksz threshold as a function of frequency below 30 kHz.

At 5 °C above the transition of SmZ$_A$ to the apolar paranematic phase N, the steady striped state is found for frequencies above 30 kHz [Fig. 14(a)]. The critical period and wave number in this patterned state vary with frequency in the same manner as in the SmZ$_A$ phase (cf. Fig. 7), except that the wave number does not tend to saturate. Between 1 kHz and 30 kHz, only the uniform Freedericksz state is obtained [see inset, Fig. 14(a)]. Brochard-Leger walls [43,44], appear transiently at lower frequencies; collapsing elliptical walls separating regions of opposite tilts indicate the ratio of splay to twist elastic constants ($k_{33}/k_{22}$) to be ~3; see Fig. 15. At higher frequencies, the walls exist as dynamically stable defects.

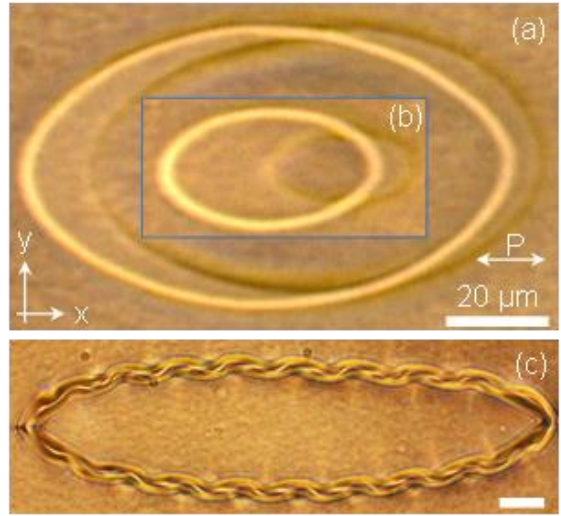

FIG. 15. Brochard-Leger loop walls in the N phase of DIO at 85 °C. (a, b) A cylindrical domain slowly collapsing after turning the field off from 1 V, 10 kHz appearing sheared along x, the rubbing direction. The two frames are about 70 s apart. From the ratio of principal axes a/b, $k_{33}/k_{22}$=(a/b)$^2$≈3. (c) Closed domain wall with a complex structure (corkscrew instability) resulting in a wavy focal line in the field on state with $U$=2 V, $f$=10 kHz; the loop is extending slowly and mainly along x.

Some representative patterned states observed under increasing voltage at 50 kHz are shown in Fig. 16. Panels (a)-(c) therein correspond, respectively, to the classical normal, oblique and bimodal roll states of electroconvection. The 2-dimensional instability in Panel (c) turns chaotic at 1.3 V and is invaded by another vertical stripe state; see Panel (d). With further rise in $U$, the stripes first become undulatory and then undergo changes that result in a 2-dimensional lattice of focal points, as in Panel (f). At much higher frequencies [see Panels (g)-(i)], the stripes develop into chevrons as $U$ increases. These textural modifications, common to usual nematics, indicate that 50 kHz is probably close to the dielectric relaxation frequency of $\varepsilon_\parallel$, the dielectric constant parallel to the director, and the dielectric anisotropy $\Delta\varepsilon=\varepsilon_\parallel-\varepsilon_\perp$ has a small positive value [12,17]. This would mean that the dielectric loss ($\varepsilon_\parallel''$) contributes to the net conductivity and the periodic zig-zag pattern can result from the EHD instability of the sample in the conduction regime, though the director has to oscillate with the field to contribute to the conductivity due to dielectric loss.

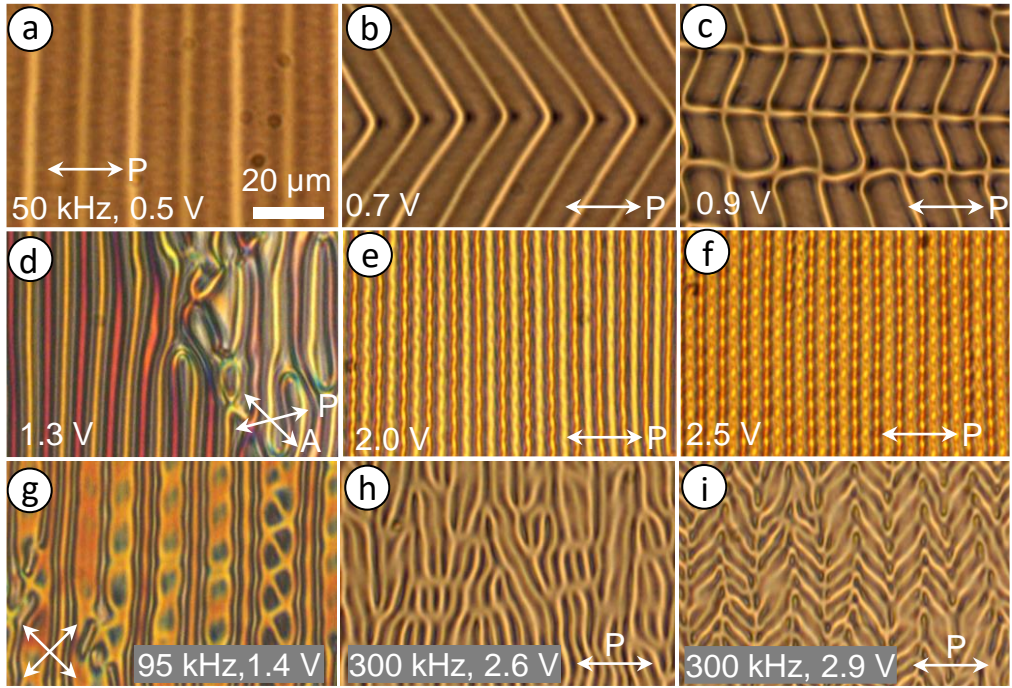

FIG. 16. Patterned states formed under increasing constraint in the N phase of DIO at $T_r$=5 ºC.

## C. Electrohydrodynamic instability in the ferronematic phase

Periodic patterns associated with a great deal of fluid motion can be readily excited in the $N_F$ phase. Two significant aspects of this instability are the sustained pattern drift and inplane vortical flows evidencing electroconvection. Figure 17 shows the general appearance of the modulated phase at low voltages. It is clear from Fig. 17(a) that the pattern wave vector is not correlated with the rubbing direction. Secondly the wavelength $\lambda$, which is the separation between the adjacent brighter focal lines, with feebler lines of the opposite focal plane present in-between, is 3-5 times the sample thickness over a wide frequency range. In a study of a 9 μm thick sample, we found $\lambda$ to remain constant at 30 μm over 10 kHz to 140 kHz, at voltages slightly above threshold, while during early evolutionary stage very close to threshold, it was around 50 μm. In Figs. 17(b) and 17(c) showing relative shift of the domains along the wave vector, we demonstrate the travelling wave nature of the instability. The stripes do not always appear over the entire sample area. For example, in Fig. 17(d), showing a common $y$-geometry of ferroelectric domains with three walls or disclination branches, the instability has appeared manifestly at lower left and is hardly discernible elsewhere.

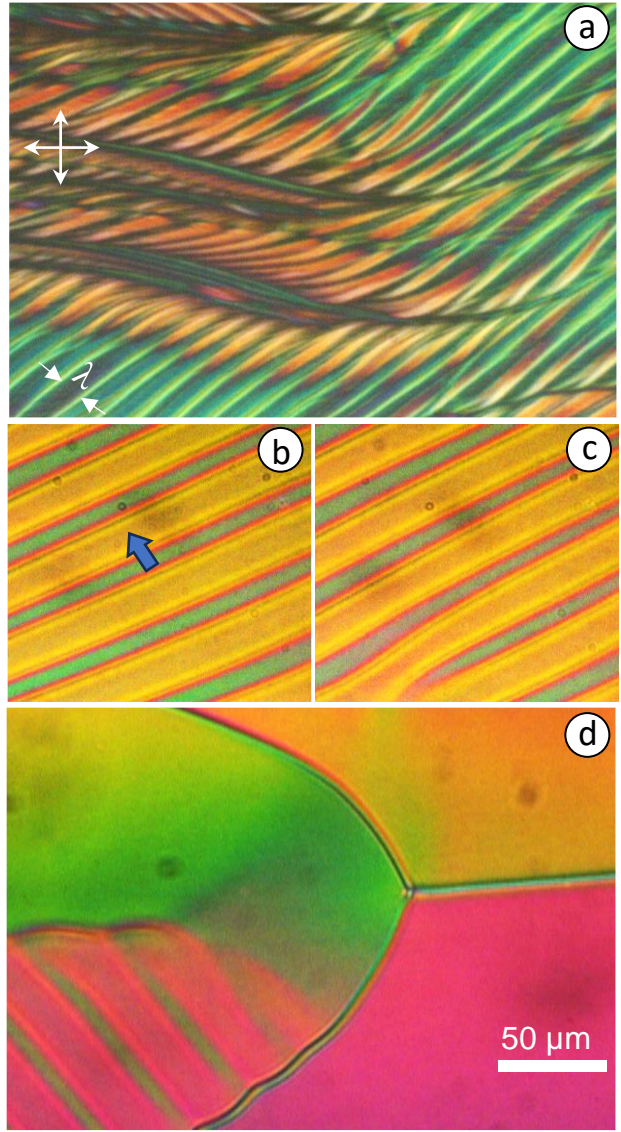

FIG.17. Drifting striped patterns in the $N_F$ phase of DIO. (a) Stripes at varying angles relative to the rubbing direction in different regions of a 5 μm thick sample, between crossed polarizers P(0)–A(90); 1.83 V, 35 kHz, 61.5 ℃. (b,c) Two frames separated by 22 s from a time series showing diagonal domains in a 9 μm thick sample; 2.4 V, 30 kHz, 62 ℃; the drift of the pattern along the wave vector is evident from a comparison of the two images. (d)Three different regions of continuously varying director fields separated by walls and disclinations in a 9 μm thick sample at 61.5 ℃ subjected to a 2.9 V, 140 kHz field.; the threshold for the instability varies for the three regions. The scale bar is common to all the Panels.

On increasing $U$, the domains begin to develop extensively. Quite frequently, parallel domains in adjacent regions separated by a defect line tend to drift in opposite directions. As for the streamlines of flow, they lie along the stripe direction. Foreign particles on the two sides of focal lines are often seen to travel in opposite directions. The vortical nature of flow is indicated by foreign particles making U-turns at edge dislocations.

## IV. DISCUSSION

The above description of instabilities leads to a plausible physical mechanism for the occurrence of periodic modulation of the planar base state with $\mathbf{q} \parallel \mathbf{n}_o$ in the SmZ$_A$ phase at relatively low voltages. At these voltages, it is very likely that the medium responds to the electric field as an apolar nematic, coupling to $\mathbf{E}$ through dielectric anisotropy [12,45]. There are a few measurements of the frequency dependences of the dielectric constants of DIO [3, 46,47]. In their original paper Nishikawa et. al. [3] have reported that both $\varepsilon_\parallel$ and $\varepsilon_\perp$ have values $\approx 400$ between $f = 100$ and $10^5$ Hz at 85 ºC, just above the SmZ$_A$ to N transition point. In the SmZ$_A$ phase *both* $\varepsilon_\parallel$ and $\varepsilon_\perp$ decrease reflecting the antiferroelectric character of the phase, and have values below 100 at 75 ºC. The imaginary parts of both components exhibit broad maxima around $10^3$ Hz at the two temperatures. These and later dielectric measurements [46,47] and our own measurements on $\varepsilon_\perp$ in Fig. 18 are in broad agreement with the trends noted above. For example, in Fig. 18(a), $\varepsilon_\perp$ at 1 kHz is about 72 compared to the corresponding value of 85 in [3]. Similarly, the low frequency relaxation mode of $\varepsilon_\perp$ in the kHz region around 75 ºC in Fig. 18(b) is one of the modes also reported in [47,48]. Additionally, as observed in [48], the low frequency relaxation of $\varepsilon_{\parallel}$ is again located in the kHz region. These relaxations will contribute to the effective conductivity (through the term $2\pi f \varepsilon$") and render the electroconvective behaviour akin to that in ordinary nematics around the dielectric inversion point. The sudden increase in the conductivity $\sigma$ at 1 kHz in Fig. 18(a) is illustrative of this aspect. The rapid decrease in $\Delta\varepsilon$ with increasing $f$ as inferable from the inset in Fig. 14(a) is also a pointer to the possible low frequency relaxation of $\varepsilon_{\parallel}$. We are thus led to the view that both the zig-zag periodic instability observed in the N phase as well as the periodic structure with $\mathbf{q} \parallel \mathbf{n_o}$ in the SmZ$_A$ phase at relatively low voltages are EHD patterns with the conductivity of the medium enhanced by dielectric relaxation processes. The zig-zag pattern in the N phase arises from the nonzero azimuthal distortion caused by a low value of the twist elastic constant (as was found from the aspect ratio of the elliptic wall in Fig. 15) and flexoelectric contribution to EHD [25,26,48].

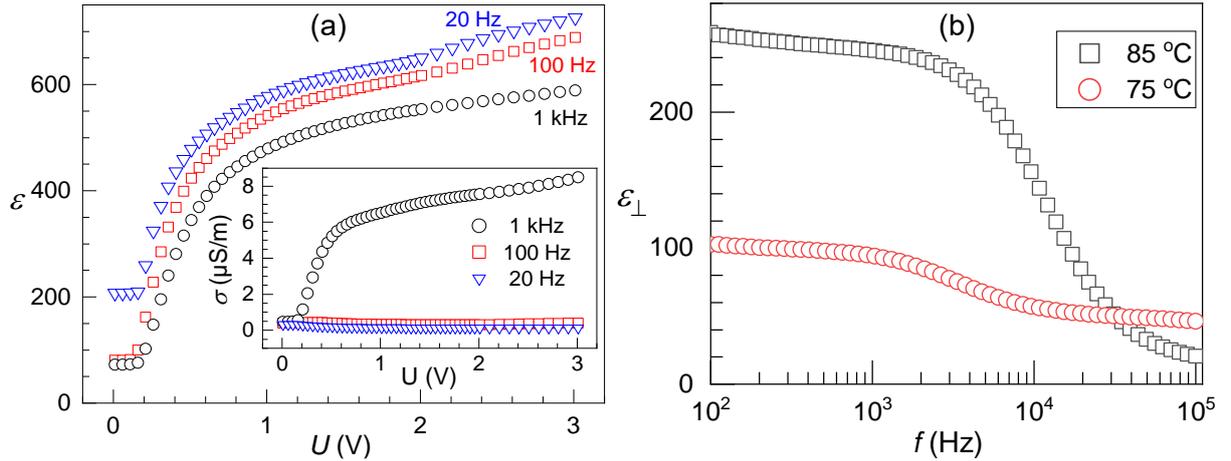

FIG. 18. Voltage variation of (a) dielectric permittivity and (inset, a) electrical conductivity at different frequencies in an initially planar, 5 μm thick layer of DIO at 75 ºC; the permittivity close to 0 V is $\varepsilon_\perp$. (b) Low frequency relaxation of $\varepsilon_\perp$ in a 20 μm thick planar layer of DIO; 10 mV field.

As we noted earlier, though non-zero $\varphi$ distortion has been found experimentally (Fig. 3) in the SmZ$_A$ phase, $\mathbf{q}$ is along $\mathbf{n_o}$. The periodic structure in this phase develops in the book-shelf arrangement of layers with a spacing of $\sim$10 nm. The surface anchoring of $\mathbf{n_o}$ of the antiferroelectric medium with such a small layer spacing can be expected to be *strong* at relatively small values of the voltage. Any $\varphi$ distortion in the normal stripes under the field will result in a change in the layer spacing, which is energetically expensive. On the other hand, the experimental studies of Chen et al. [6] have shown that, as the temperature is reduced, in a field free sample with book shelf structure, a small reduction in the layer spacing leads to a tilting of the layers and a chevron structure enforced by the strong surface anchoring. In this case, the separation between the layers in planes parallel to the surface is equal to that at the strongly anchored surface, which itself is fixed to the high value corresponding to that at a higher temperature, as in the case of a SmC sample cooled from the SmA phase. We propose that in our experiments, the layers develop a sinusoidal *undulation* along $\mathbf{n_0}$ to fill space (Fig. 19). As the lateral boundaries of the sample are essentially free, there is no compression of the layers, i.e., the layer spacing itself is unaltered, and the only energy cost of the undulation arises from the additional curvature distortion of the **n**-field which follows the undulations of the layers. It is natural to assume that the wavelength of the undulation equals that of the periodic instability, and the resulting structure has a finite spatial oscillation of the azimuthal angle $\varphi$ as seen in experiments reported above.

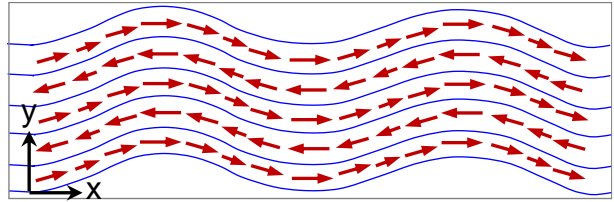

FIG. 19. A simple model for the striped state formed at the primary bifurcation of a planar SmZ$_A$ slab. The schematic diagram shows the top view of the polarization distribution in the undulating layers above the threshold voltage of the periodic instability. The layer thickness ($\sim$10 nm) is highly exaggerated in the figure compared to the stripe or undulation wavelength ($\sim$ a few $\mu$m). Arrows indicate the polarization field, with their projected length depending on the local tilt. Note that at the low voltages at which this periodic instability is observed, the $\mathbf{P}$ vector in neighbouring layers remains antiferroelectric, and the response to the electric field is similar to that of the higher temperature apolar nematic. The tilt of $\mathbf{n}$ is a result of the viscous torque due to the upwelling (say) of the nematic at the crests, and downwelling at the troughs of the undulations.

As the voltage is increased, at lower frequencies the periodic structure gives way to a uniform field of view with a reduced optical phase difference. This probably signals that the **P** vector which is antiferroelectric in adjacent layers itself starts responding to **E** by a *linear* coupling. The polar angle will be the same in both layers when measured with respect to **E**, but with a relative scissor mode with respect to each other. In that case it is clear that the medium cannot have a periodic structure seen at lower voltages at which the response is quadratic in **E**. As the voltage is increased further, the time average value of $<\!\!-\mathbf{P}\cdot\mathbf{E}\!\!>$ is large enough to completely overcome the strong interlayer antiferroelectric electrostatic interaction, and the medium is essentially in the field induced ferroelectric nematic phase. The layer structure is lost and **P** of the medium responds to the oscillatory **E** field. This can in turn give rise to oscillatory ionic charges which are screened by the polarization charges. The orientational order required by **P** implies that the oscillatory motion of **P** leads to a corresponding flow in the medium. This appears to be the origin of the second periodic instability above the higher threshold voltage which grows from the uniform state. As the layers are absent, the stripes can support $\varphi$ distortion of the **P**-field, resulting in an oblique angle between **q** and $\mathbf{n_0}$; see Fig. 10(a). As the space and time averaged orientation of **P** continues to be in the plane parallel to the electrodes, there is an intrinsic polar orientation in that plane, and the flows generated can have an average component along that vector, giving rise to the travelling waves seen in the medium. Another contributory factor can be a significant difference in the mobilities of positive ions and negative ions [49,50]. As the magnitude of polarization decreases on increasing the temperature in the SmZ$_A$ phase [3], the threshold for the TW instability goes up rapidly with temperature; see Fig. 12. The velocity of TW also increases both due to increased $U$ and lowered viscosity. At 75 ºC, the stripes become straight and no longer travel if the frequency is increased to 50 kHz. One possibility for this change is that the large viscous dissipation decreases the amplitude of **P** reorientations. The ions also cannot follow the rapid oscillations of **E**, and essentially become static. As the voltage is increased, the bright focal lines of the striped state are modulated into arrays of intense focal spots along their length, giving rise to a two-dimensional structure. Eventually a hexagonal arrangement of bright spots, slightly elongated in a direction orthogonal to $\mathbf{n_o}$ is observed. At present we can only speculate that this arrangement may minimise the electrostatic energy of the ions and the oscillating $\mathbf{P}$-field.

It is relevant to mention here of a very recent report by Basnet et al [51] on the effect of high frequency (200 kHz) electric field in planar aligned samples of RM 734 (Fig. 1), in both the nematic and ferroelectric nematic phases. The usual splay Freedericksz transition is found in the N phase, with $K_{33}/K_{11}$ estimated to be ~ 5 to 6. In the $N_F$ phase the linear coupling between $\mathbf{E}$ and $\mathbf{P}$ results in oscillations of $\mathbf{P}$ at the frequency of the applied field. Beyond a threshold voltage, the $\mathbf{P}$ field exhibits a periodic structure with both splay-bend and splay-twist distortions, such that the net splay gets cancelled and bound charges are not found. At a higher threshold voltage, a pure splay-bend distortion of the $\mathbf{P}$ field is found, resulting in a two-dimensional lattice of ±1 defects. The splay cancellation is not complete and the medium exhibits some hydrodynamic flow as well. Interestingly, both the 1D and 2D structures are attributed to a competition between *dielectric* and elastic energies, the amplitude of oscillation of $\mathbf{P}$ limited to small values by the strong viscous response of the medium. Thus, the electric field response of RM 734 in the $N_F$ phase appears to be quite different from that of DIO which also exhibits the intermediate $SmZ_A$ phase.

## V. CONCLUSIONS

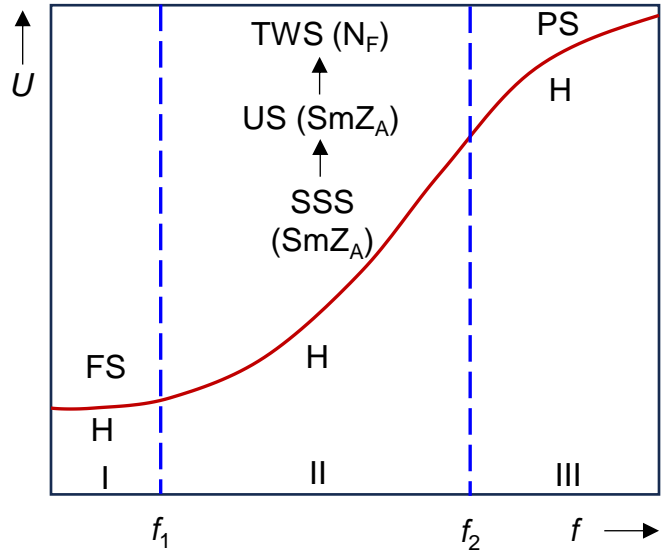

FIG. 20. Phase diagram of DIO showing qualitatively the instabilities at 75 °C. In region I below frequency $f_1$, the planar H state bifurcates into the Freedericksz state FS above a voltage $U_F$; in region II between $f_1$ and $f_2$, the primary instability is the stationary stripe state SSS; secondary bifurcations into uniform and travelling wave states (US and TWS) follow at higher voltages. In region III for $f > f_2$, 1- and 2-dimensional patterned states (PS) occur as in the dielectric relaxation region of ordinary nematics.

We have presented experimental studies on the pattern forming instabilities under the action of AC electric fields on different liquid crystalline phases exhibited by planar aligned DIO, which exhibits the phase sequence Nematic-$SmZ_A$-Ferroelectric Nematic as it is cooled from the isotropic phase [3]. The nematic has a positive $\Delta\varepsilon$ at low frequencies, and only a dielectric reorientation above a threshold voltage is seen, as usual. As the frequency is increased above a critical $f$ (for example, above 30 kHz at 5 ºC above $T_{AF}$), the planar cell exhibits a zig-zag EHD instability, which owes its origin to the approach to a relaxation frequency of $\varepsilon_\parallel$, and the resulting enhancement in the effective conductivity of the sample [13]. In the SmZ$_A$ phase, at low frequencies (100 Hz-4 kHz), the usual dielectric reorientation involving a quadratic coupling between **E** and $\Delta\varepsilon$ is seen as the primary instability, and unspecific chevron-like structures form randomly as secondary effects; a robust periodic state is obtained only above ~5 kHz at 75 ºC. Its wave vector **q** remains parallel to **n$_o$** despite the presence of an azimuthal distortion of the director field, that changes sign between adjacent stripes. We argue that the latter distortion arises from an undulation instability of the layered structure. The threshold voltage increases quadratically, and **q** linearly with the frequency $f$ (upto~200kHz). These results are consistent with the expectations for EHD instability around the inversion frequency of $\Delta\varepsilon$ [15]. As the voltage is increased, the time averaged energy of the linear coupling between **E** and **P** can exceed the electrostatic interaction between the layers giving rise to the antiferroelectric order at zero field. The scissoring type reorientation between neighbouring layers under the oscillating electric field is geometrically incompatible with the viscous flow giving rise to EHD, and the medium goes over to an optically uniform state, with a path difference which is lower than that in the field free sample. At a higher threshold voltage, the average electric energy is large enough to completely overcome the remnant antiferroelectric interaction between the layers, and the medium goes over to a field induced ferroelectric phase. The layering is lost, and the uniform **P** now oscillates with the field. The *time averaged* orientation of **P** is parallel to the plates. The average **P** field can have bend fluctuations, which can in turn generate charge separation which is essentially static, and an EHD instability analogous to the 'dielectric regime' studied in apolar nematics. The *polar* nature of the medium can, however, give rise to a TW structure, seen in a range of frequencies. As the frequency and voltage are increased, the TW structure progressively transforms into a two dimensionally modulated dynamical structure. We can only speculate that it probably arises from a reduction in the amplitude of **P** oscillations due to viscous damping, and breakup of the sheets of ionic charge densities in the stripes due to their repulsive interactions.

## ACKNOWLEDGMENTS

The authors KSK, SKP and DSS are thankful to Prof. B. L. V. Prasad, Director, Centre for Nano ad Soft Matter Sciences for experimental facilities and useful discussions. RJM acknowledges funding from UKRI via a future leaders' fellowship, Grant number MR/W006391/1.